# SELM: Software Engineering of Machine Learning Models

Nafiseh Jafari, ICT Department, Malek Ashtar University of Technology, Tehran, Iran

Mohammad Reza Besharati[1], Department of Computer Engineering, Sharif University of Technology, Tehran, Iran

Mohammad Izadi, Department of Computer Engineering, Sharif University of Technology, Tehran, Iran

Maryam Hourali, ICT Department, Malek Ashtar University of Technology, Tehran, Iran

**Abstract**

One of the pillars of any machine learning model is its concepts. Using software engineering, we can engineer these concepts and then develop and expand them. In this article, we present a SELM framework for **S**oftware **E**ngineering of machine **L**earning **M**odels. We then evaluate this framework through a case study. Using the SELM framework, we can improve a machine learning process efficiency and provide more accuracy in learning with less processing hardware resources and a smaller training dataset. This issue highlights the importance of an interdisciplinary approach to machine learning. Therefore, in this article, we have provided interdisciplinary teams' proposals for machine learning.

## 1. Introduction

Machine learning usually aims to find and develop a computational model for an intelligent task on a practical problem. Development based on calculation can be called engineering (1). In software engineering, software systems are calculated and engineered by models. In fact, these software models are platforms for analysis, design, development, and system engineering.

---

[1] besharati@ce.sharif.edu

For modeling in software engineering, we need several elements: 1. The modeling perspective, 2. The system under modeling, and 3. Modeling language and tools (5). So we look at a system under modeling from one or several perspectives, and we discover a set of meanings about that system. Using the modeling language and tool, we express and record those perceived meanings of the system. This allows us to engineer the system under modeling (as-is system or to-be system) by changing and transforming it.

We could perhaps consider the modeling perspective as the most crucial part of this conceptual architecture. Because if we do not use a proper perspective or perspectives to look at the meanings of a system, it is practically impossible for us to achieve a model, start modeling, and ultimately develop through engineering calculations. To obtain the proper perspective or perspectives of a system, we need human intuition of that system, meaning there is a deep connection between the modeler's intuition and system engineering (including software, hardware, and intelligence). Therefore, we can consider human intuition as the infrastructure for system engineering activities.

## 2. Reviewing the methods of using human intuition in machine learning and intelligence

*Method 1:* Metavariable (11), hyperparameter (9), and architectures (10).

*Method 2*: Labeling, supervision, selection, pruning, and data engineering (12).

*Method 3:* Feature selection, feature definition, feature weighting (12, 22, 23).

*Method 4*: Fuzzification and fuzzy rules (14, 15).

*Method 5*: Models based on knowledge representation, such as ontologies, frames (16), description logics (17), formal notations (18), the semantics of logic (3), and so forth.

*Method 6*: Mass input gathering of human data (19).

*Method 7:* Mass input gathering of the learning experience in the environment with human presence (20).

*Method 8*: Utilizing human regulations, filters, pre-processing, post-processing, and normalization.

*Method 9:* Utilizing heuristic algorithms.

*Method 10*: Informatics engineering of data by humans before machine processing, data structuring, classifications, modeling identities, taxonomies, and so forth.

*Method 11*: Designing a basic model and learning platform using human intuition (e.g. a variety of trees in NLP).

*Method 12*: Basic axiomatic systems, such as grammars, logics, axioms, and so forth.

*Method 13*: Basic models (e.g. age, Markov, etc.), reference models, domain models, performance models, environment models (4), implementation models, analytical models, meta-level models, and viewing angles.

*Method 14*: Surveys, counting, statistical distributions, probability models, statistical models, and PGM.

*Method 15* (duplicated in previous cases): Databases, information bases, knowledge bases, logs, datasets, value catalogs (e.g. color or cognitive catalogs).

*Method 16*: Computing and calculating methods, basic computational models (wavelet analysis, Lisp, Turing machine, regression, graph traversal and possible worlds, topologic functions, set theory, type theory, Rio, etc.)

*Method 17*: Theories

*Method 18*: Reutilization of previously learned models in new fields and transfer learning.

*Method 19:* Description of a simulation system by a human to simulate another system (2).

*Method 20:* Transformation of semantic qualities into computational quantities through mapping logics defined by human expertise (7).

*Method 21:* Imitating the cognitive processes of intelligent living agents, such as humans and animals. In this way, they study various cognitive mechanisms in the brain, mind (21), and cognitive behavior of living intelligent agents and try to achieve a formulation and reverse engineering to explain how these processes work. One of the most famous examples of this approach is neural reinforcement learning (8).

## 3. The proposed framework

Figure 1 shows the process life cycle of the SELM framework. With the help of this process life cycle, which should be performed gradually, incrementally, and repetitively in several cycles, we can software engineer the machine learning models.

In this life cycle, both the order of execution of affairs and the chain of artifacts resulting from the execution of these affairs are included.

The foundation and starting point of this process is human intuition. We can use all the methods listed in the previous section to provide and apply human intuition for the machine learning cycle. To integrate these models, especially to integrate the solution and problem models, we can use the semantics of logic, described in detail in (1) and (7).

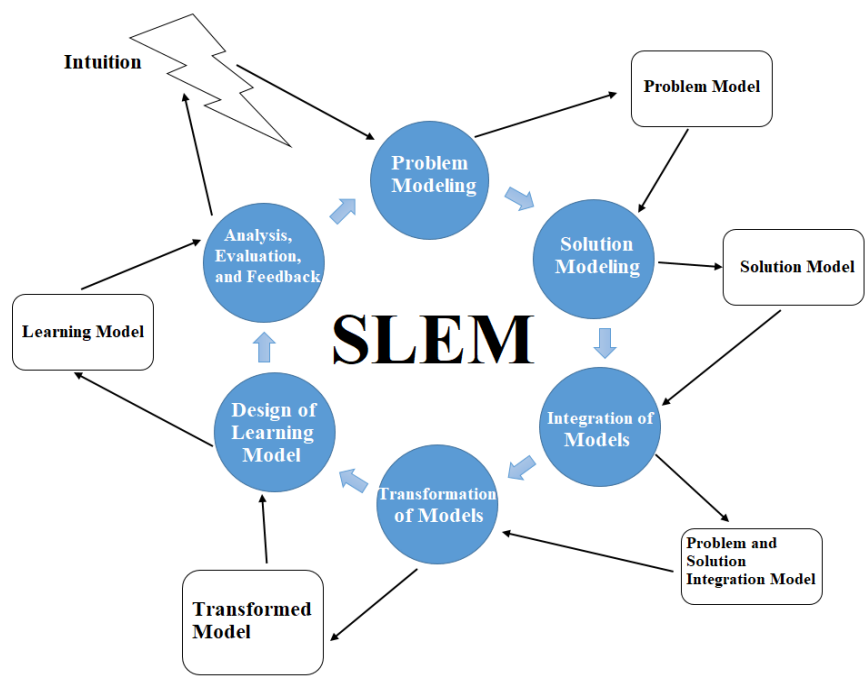

**Figure 1.** *The process life cycle of the SELM framework*

## 4. Case study

Using the IR-QUMA data set (6), which is about the mobile messaging software quality from the perspective of Iranian users (1), we conducted a case study to examine the effectiveness of the SELM framework. We wanted to know if it was possible to achieve higher efficiency in the machine learning process by applying the SELM framework activities. To evaluate machine learning models, we used the 10-Fold Cross-Validation method on 2837 data records, and we also used a Running Average window with a length of 20 as the input filter. Table 1 shows the results of this case study. The results show that with the same volume of learning data and the same learning algorithm, using the SELM framework has improved the learning accuracy by better engineering the features and more efficient use of background knowledge.

Table 1. *Results of case study*

| Name of the classification algorithm | Accuracy percentage when applying SELM | Accuracy percentage without applying SELM | Rate of change |
|---|---|---|---|
| Random Forest | 94.55 | 93.34 | +1.22 |
| Multilayer Perceptron | 88.54 | 86.85 | +1.69 |

## 5. Conclusion

To achieve higher efficiency levels in the machine learning process, we can synergistically use the knowledge, experience, and results of other specialized fields (especially software engineering). We can only achieve this through the integration of engineering processes and the provision of interdisciplinary teams (which include both data engineers and scientists, as well as other engineers and scientists in software and computing sciences). We can see examples of this in the experience of other countries and leading companies in the field of cognitive intelligence (specifically, IBM at the Watson Research Center and on the Watson Machine Project, in which for about five years the company gathered a team of about 50 people from about 30 different specialties in various fields of science and engineering, such as logic and philosophy, mathematics and statistics, data and software engineering, control and algorithm, neuroscience and biology, and praxeology and psychology).